\documentclass[prl,aps,superscriptaddress,amsmath,amssymb,twocolumn]{revtex4-1}
\usepackage{amsmath,amssymb}
\usepackage{graphicx}
\usepackage{bm}
\usepackage{xcolor}

\usepackage[colorlinks=true,linkcolor=blue,citecolor=blue,anchorcolor=blue,urlcolor=blue,pdfborderstyle={}]{hyperref}

\newcommand{\blue}{ }
\newcommand{\magenta}{ }
\newcommand{\bea}{\begin{eqnarray}}
\newcommand{\eea}{\end{eqnarray}}
\newcommand{\be}{\begin{equation}}
\newcommand{\ee}{\end{equation}}
\def\Ai{\text{Ai}}

\def\Ai{\text{Ai}}

\DeclareMathOperator{\cov}{cov}
\newcommand{\cum}[2]{\langle {#1} \rangle^\mathrm{c}_{#2}}
\newcommand{\pref}[1]{(\ref{#1})}

\newcommand{\nn}{{\nonumber}}
\def\be{\begin{equation}}
\def\ee{\end{equation}}

\def\l{\lambda}

\def\l{\lambda}

\begin{document}
\title{Memory and universality in interface growth}

\author{Jacopo De Nardis}
\email{jacopo.de.nardis@phys.ens.fr}
\affiliation{D\'epartement de Physique, Ecole Normale Sup\'erieure, PSL Research University, CNRS, 24 rue Lhomond, 75005 Paris, France}
\author{Pierre Le Doussal}
\email{ledou@lpt.ens.fr}
\affiliation{CNRS-LPTENS, Ecole Normale Sup\'erieure, PSL Research University, 
24 rue Lhomond, 75005 Paris, France.}
\author{Kazumasa A. Takeuchi}
\email{kat@kaztake.org}
\affiliation{Department of Physics, Tokyo Institute of Technology, 2-12-1 Ookayama, Meguro-ku, Tokyo 152-8551, Japan.}
\date{\today}


\begin{abstract}

Recently, very robust
universal properties were shown to arise in one-dimensional growth processes
with local stochastic rules,
leading to the 
Kardar-Parisi-Zhang universality class.
Yet 
it has remained essentially unknown
how fluctuations in these systems correlate at different times.
Here we derive quantitative predictions for the universal form of the 
two-time aging dynamics of growing interfaces {\blue and we show from first principles the breaking of ergodicity that the KPZ time evolution exhibits. }
We provide corroborating experimental observations 
on a turbulent liquid crystal system, as well as a numerical simulation of the Eden model,
{\blue and we demonstrate the universality of our predictions}. These results may give insight
into memory effects in a broader class of 
far-from-equilibrium systems.
\end{abstract}
\maketitle


%


\paragraph*{Introduction.} Non-equilibrium dynamics is ubiquitous in nature, and takes diverse forms, such as
avalanche motion in magnets and vortex lines \cite{Aging,AvalanchesShape} 
ultraslow relaxation in glasses \cite{Kurchan,Leticia}, 
unitary evolution towards
thermalization in isolated quantum systems \cite{Eisert}, { coarsening
in phase ordering kinetics \cite{coarsening},} and
flocking in living 
matter \cite{toner}. Prominent examples 
are growth phenomena, which abound in physics \cite{StanleyBook,TS2010,TS2012,combustion},
 biology \cite{StanleyBook,Wakita1997,nelson}, and beyond \cite{Atis2015}.
As { some of these} systems try to reach local equilibrium or
stationarity, a great variety of behaviors can occur,
such as aging dynamics and memory of past evolution \cite{Leticia,Aging,coarsening,Henkel}.
{ How universal and generic are these behaviors is
a fundamental question \cite{Henkel}.}

One important example of growth 
arises when a stable phase of a generic system expands into a non-stable (or meta-stable) one, { in presence of noise.} While spreading, the interface separating the two phases develops many non-trivial geometric and statistical features. A universal 
behavior then emerges, 
unifying many growth phenomena into a few universality classes, irrespective of their microscopic details. The most generic one, 
for local growth rules,
is the celebrated Kardar-Parisi-Zhang (KPZ) class, now substantiated by many experimental examples, such as growing turbulence of liquid crystal \cite{TS2010,TSSS2011,TS2012}, propagating chemical fronts \cite{Atis2015}, paper combustion \cite{combustion} and bacteria colony growth \cite{Wakita1997}. 
For one-dimensional interfaces growing in a plane,
 as studied in many experiments,
it is characterized by the following KPZ equation \cite{KPZ}:
\be \label{kpzeq}
\partial_t h(x,t) = \nu \partial_x^2 h(x,t) + \frac{\lambda_0}{2}  (\partial_x h(x,t))^2 + \sqrt{D} ~ \xi(x,t) 
\ee
{ which describes} the motion of an interface of height $h(x,t)$ at point $x \in \mathbb{R}$
at time $t$, 
driven by a unit space-time white noise $\xi(x,t)$. 
Recently, this 
problem became an outstanding example where a wealth of universal statistical properties can be solved exactly, from
the KPZ equation and related lattice models 
\cite{
png,spohn2000,spohnKPZEdge,we-narrow,dotsenko,corwinDP,KPZFlat,KPZStat,largefluct1,largefluct2,crissingprob2}.
\begin{figure}[t]
 \centering
 \includegraphics[width=0.95\hsize, height=6.9cm,keepaspectratio]{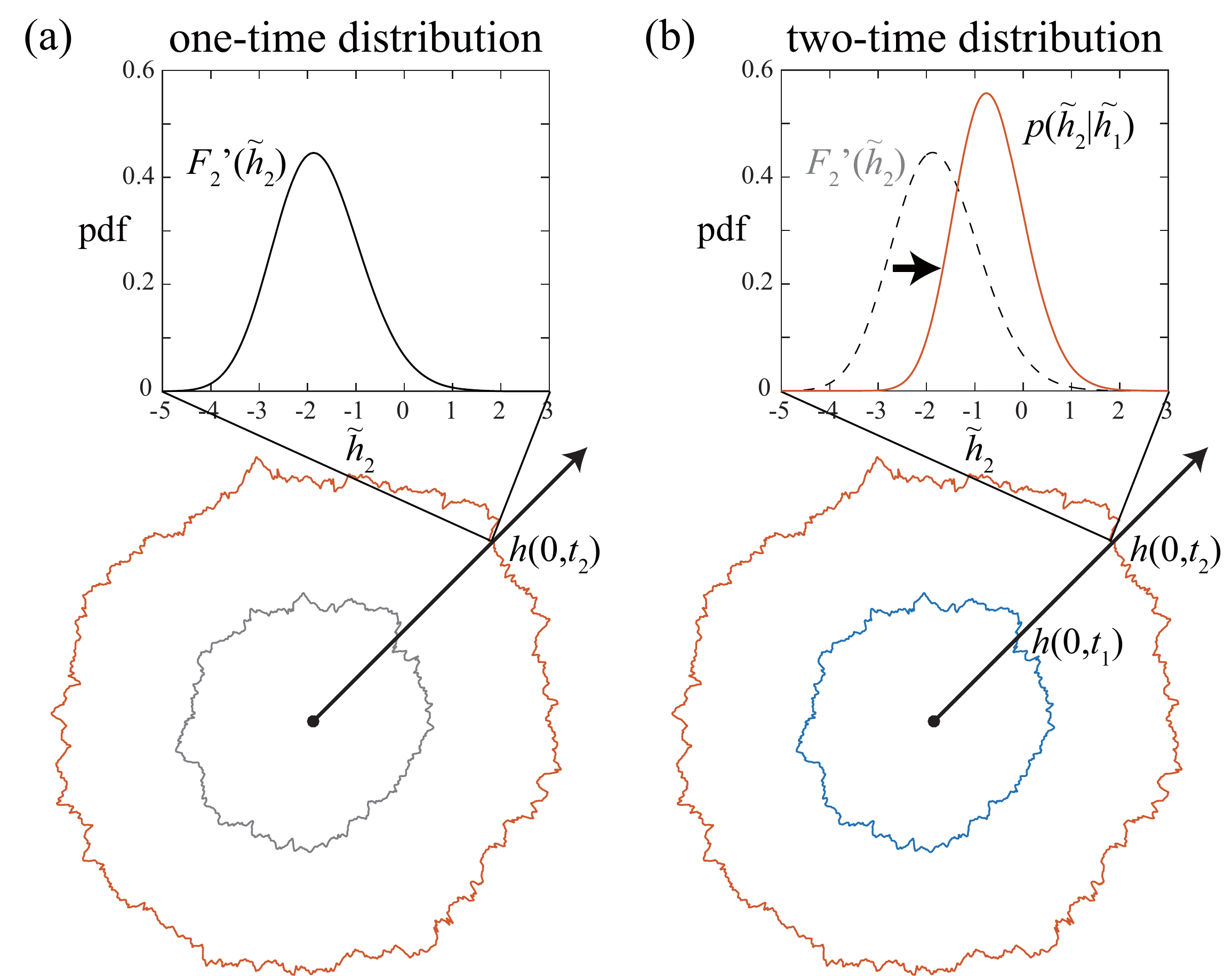}
  \caption{ 
{\magenta Sketch of the KPZ time evolution}: the two rough lines show the expanding KPZ height, {\magenta describing the edge of a growing circular region. Note that, for isotropic systems, the local radius corresponds to $h(0,t)$ in any angular direction.}
(a) The fluctuations of the interface at time $t_2$ are described by the GUE Tracy-Widom distribution, $F_2'(\tilde{h}_2)$, with the rescaled height $\tilde{h}_2 := \tilde{h}_{t_2} = \frac{h(0,t_2) - v_\infty t_2}{(\Gamma t_2)^{1/3}}$.
(b) Given the fluctuations of the height at a previous time $t_1 = t_2(1 + \Delta)^{-1}$ along the same angular direction (black arrow), the two-time conditional probability density $p(\tilde{h}_2|\tilde{h}_1)$ (red line) measures the probability of observing a fluctuation {\magenta value} $\tilde{h}_2$ at time $t_2$, given {\magenta the value of} $\tilde{h}_1$ at the previous time $t_1$ {\magenta (the inset shows the conditional distribution for} $\tilde h_1=0$). {\magenta This therefore quantifies} how much memory of the previous height configuration is kept during the time evolution.
}
  \label{fig-sketch}
\end{figure}%
{\blue
At large time, the height 
evolves as 
\begin{equation}
h(0,t) \simeq  v_{\infty} t  + (\Gamma t)^{1/3} \small{\tilde{h}_t}
\end{equation}}%
with system-dependent parameters $v_{\infty},\Gamma$ and a stochastic variable $\tilde{h}_t$ that carries universal information of the fluctuations. Remarkably, in the limit $t \to \infty$, $\widetilde{h}_t$ follows one of a few non-Gaussian universal distributions, selected only by the global geometric shape of the initial condition $h(x,t=0)$: in particular, the GUE Tracy-Widom distribution \cite{TW1994}, $F_2(\sigma)$, when $h(x,0)$ is narrowly curved [droplet initial condition
\cite{spohnKPZEdge,we-narrow,dotsenko,corwinDP}, see Fig.~\ref{fig-sketch}(a)] 
and its GOE variant, $F_1(\sigma)$, 
when $h(x,0)$ is a flat surface \cite{KPZFlat}.
These two distributions also describe the fluctuations of the largest eigenvalue of a gaussian random matrix
drawn 
from the unitary (GUE) or orthogonal (GOE) ensembles, revealing a striking 
connection to the theory of random matrices \cite{spohn2000,KrugReview}. 
An additional universal distribution,
 the Baik-Rains distribution \cite{png}, 
characterizes the stationary 
state 
of the growth 
and 
can be reached \cite{KPZStat} by choosing $h(x,0)$ as 
Brownian motion in $x$.

This geometry-dependent universality was 
tested and confirmed experimentally, in 
studies on growing interfaces of liquid-crystal turbulence
 \cite{TS2010,TSSS2011,TS2012}.
The experiments also allowed to investigate
 time-correlation properties
that were inaccessible by analytical approaches. This revealed an anomalous memory effect for the droplet case \cite{TS2012}, { by which} fluctuations in $h$ keep indefinite memory of the past, in contrast to the naive expectation that memory is eventually lost.
This persistence of memory,
 signaling ergodicity breaking in the time evolution of the droplet case,
 is quantified by the long time limit of the covariance {\blue that remains strictly positive \cite{TS2012,FerrariSpohn2times}
\begin{equation}
\lim_{\Delta \to \infty}\lim_{t_1 \to \infty}C(t_1,t_1(1+ \Delta)) > 0.
\end{equation} 
where $C(t_1,t_2)  = \cov[h(0,t_1), h(0,t_2)]$. }
Theoretically, however, such two-time quantities {\blue{remained so far}} analytically intractable, except for a few exceptional results \cite{dotsenko2times1,Johansson2times} that however are too involved to produce practical predictions. {\magenta 
Since experiments and simulations are always confronted
 with relatively limited ranges of time $t_1$ and ratio $\Delta$,
 while the suspected ergodicity breaking can only be addressed
 in the limits $t_1 \to\infty$ then $\Delta \to \infty$,
 theory that can directly deal with these asymptotic limits,
 and also make a bridge to finite-time observations through predictions,
 is a crucial missing facet of the problem.}

{\blue
Here we provide first {\magenta such} theoretical results for the correlations at two different times in the infinite time limit of the KPZ equation, and we analytically prove the persistence of correlations that was previously observed in finite-time experiments 
\cite{TS2012}. This shows that a fraction of the fluctuations of the KPZ interface, namely the ones with large and positive rescaled height, maintain their configurations stable during the time evolution, translating into an ergodicity breaking in all the growth processes with the droplet initial condition. \\
 }
\begin{figure}[t]
 \centering
 \includegraphics[width=\hsize]{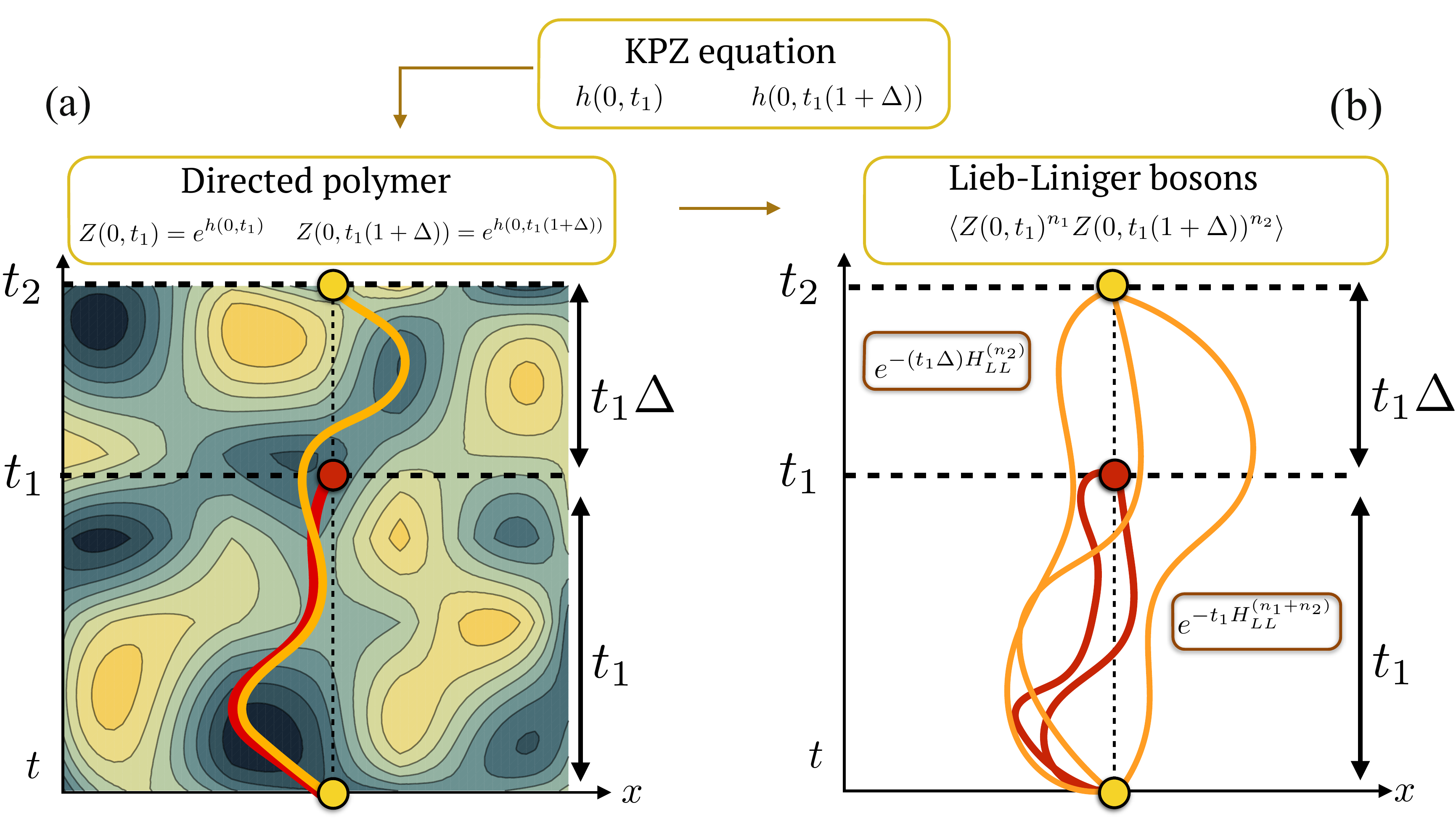}
\caption{{\blue
(a) Representation of the mapping from the height $h(x,t)$ in the KPZ equation \eqref{kpzeq} to the free energy of a directed polymer in random potential. The JPDF $P_\Delta(\sigma_1,\sigma)$ maps to the
JPDF of the free energies of two directed polymers starting both in $(0,0)$ but ending in $(0,t_1)$ and 
$(0,t_2=t_1(1+\Delta))$: shown is a typical configuration of these 2 paths, which
tend to visit the lower valleys of the potential (bluer regions), i.e. regions with large and positive fluctuations of the KPZ rescaled height $\tilde{h}_1$, compatible
with their boundary conditions and kinetic energies, which tends to minimize their length.
The two paths tend therefore to overlap in the time interval $[0,t_1]$, accounting for ergodicity
breaking (see text). In (b) is shown the mapping (via the replica trick) of the directed polymer to a quantum mechanical transition amplitude of attractive one-dimensional bosons.}}
  \label{fig-polymers}
\end{figure}%
\paragraph*{Two-time JPDF.}
{\blue 
We address the problem by deriving an analytical result for the joint probability density function (JPDF) of the height $h$ at two different times, $t_1$ and $t_2=(1+ \Delta)t_1$, with the droplet initial condition, see Fig. \ref{fig-sketch}.}
It is valid in a wide range of parameters and
 agrees remarkably well with experimental and numerical data (see below).
We focus on the limit $t_1,t_2 \to \infty$ with their ratio $t_2/t_1 = 1+ \Delta$ kept finite, so that the obtained correlations are expected to be universal within the KPZ class. More precisely, we compute the JPDF for the rescaled height
$\tilde{h}_1 := \tilde{h}_{t_1} = \frac{h(0,t_1) - v_\infty t_1}{(\Gamma t_1)^{1/3}}$ and the rescaled two-time height difference $\tilde{h}_{12} := \frac{h(0,t_2) - h(0,t_1) - v_\infty t_1 \Delta}{(\Gamma t_1 \Delta)^{1/3}}$. It is defined as
\begin{align} 
&P_{\Delta} (\sigma_1, \sigma) d\sigma_1 d\sigma \notag  \\
&\qquad = \lim_{t_1 \to \infty} {\rm Prob}\left(\sigma_1 \!\leq\! \tilde{h}_1\!\leq\! \sigma_1 \!+\! d\sigma_1 , ~\sigma \!\leq\! \tilde{h}_{12} \!\leq\! \sigma \!+\! d\sigma \right).  \label{JPDF0}
\end{align}
and quantifies 
how much memory of the configuration at the earlier time 
$t_1$ is retained at the later time $t_2$, as illustrated in Fig.~\ref{fig-sketch}(b). 
It allows to calculate the conditional cumulants $\cum{\tilde{h}_{12}^n}{\tilde{h}_1> \sigma_{1c}}$, i.e., the cumulants of the variable $\tilde{h}_{12}$
conditioned to 
realizations 
with $\tilde{h}_{1}$ larger than some fixed value $\sigma_{1c}$. 
It also allows to predict the rescaled covariance under the same conditioning, defined as
\begin{equation}
C_{\Delta,\sigma_{1c}} = \frac{C(t_1,t_2)_{\tilde{h}_1> \sigma_{1c}}}{C(t_1,t_1)_{\tilde{h}_1> \sigma_{1c}}} = 1+\Delta^{1/3} \frac{\cov[\tilde{h}_{1},\tilde{h}_{12}]_{\tilde{h}_1> \sigma_{1c}}}{\cum{\tilde{h}_{1}^2}{\tilde{h}_1> \sigma_{1c}}}.  \label{CorrDef}
\end{equation}
These quantities, computed here analytically for the first time, 
allow to probe memory effects and quantify the breaking of ergodicity in the dynamics.
{\blue In particular,
\eqref{CorrDef} quantifies how much memory of the fluctuations with rescaled amplitudes larger than $\sigma_{1c}$ is kept at later times and it recovers the full two-time covariance $C(t_1,t_2)$ in the limit $\sigma_{1c}=-\infty$.} \\ 

\begin{figure*}[t]
 \centering
 \includegraphics[width=\hsize]{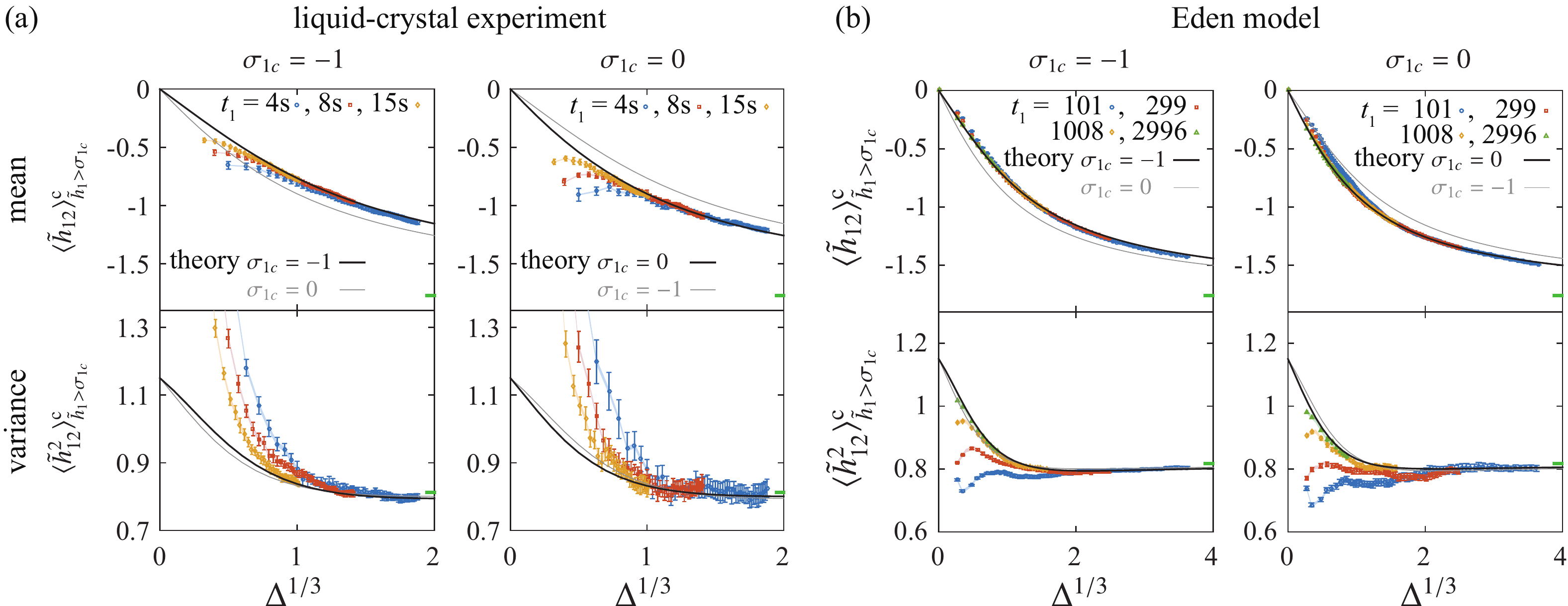}
  \caption{
Test of the theoretical prediction for the conditional mean $\cum{\tilde{h}_{12}}{\tilde{h}_1 > \sigma_{1c}}$ and variance $\cum{\tilde{h}_{12}^2}{\tilde{h}_1 > \sigma_{1c}}$ with the liquid-crystal experiment (a) and the Eden-model simulation (b). Here the results for $\sigma_{1c} = -1$ and $0$ are shown (see also Fig.~S3 in \cite{SuppMatt}for $\sigma_{1c} = -1.5$). Data at different $t_1$ are shown in different colors and symbols. The regions of overlapped data indicate the asymptotic $\Delta$-dependence, which is found to be in excellent agreement with the theoretical predictions (black lines), without any fitting parameter. For comparison, the theoretical curves with another value of $\sigma_{1c}$ are shown by gray thin lines. The error bars indicate the standard errors, and the shaded areas show the uncertainty due to the estimation error in $v_\infty$ and $\Gamma$. To reduce the effect of finite-time corrections, here we used such realizations that satisfy $\tilde{h}_1 > \tilde{h}_{1c}$ with $\text{Prob}[\tilde{h}_1 \geq \tilde{h}_{1c}] = 1-F_2(\sigma_{1c})$. The deviation of the non-overlapped data is due to finite-time corrections, which decay as $t_1^{-1}$, see Fig.~S4 in \cite{SuppMatt}). Note that the asymptotic theoretical curves converge to the Baik-Rains values (mean 0, variance 1.1504) at $\Delta \to 0$ and the GUE Tracy-Widom values (mean -1.7711, variance 0.8132, indicated in the figures by the green bars) at $\Delta\to\infty$.
}
  \label{fig-condcum}
\end{figure*}%

\paragraph*{Solution via the directed polymer.}
To derive a numerically tractable expression for the JPDF \pref{JPDF0}, 
we exploit the fact \cite{kardareplica} that the KPZ equation is equivalent to a (statistical mechanics) problem of
space-time paths (i.e. ``growth histories'') in a random potential, 
which is further mapped into a quantum problem
of bosons (see Fig.~\ref{fig-polymers}).  From now on we use the scales $x^* = \frac{(2 \nu)^3}{D \lambda_0^2}$, $t^* = \frac{ 2 (2 \nu)^5}{D^2 \lambda_0^4}$, $ h^* = \frac{2 \nu}{\lambda_0}$ as units of space, time and height, respectively.
In other words, $x/x^*, t/t^*, h/h^*$ are simply denoted by $x,t,h$, respectively (this amounts to setting $\nu=1$, $\lambda_0=2$ and $D=2$ in the KPZ equation, which leads to $\Gamma=1$). 
 In these units, from \eqref{kpzeq}, the function $Z(x,t)=e^{h(x,t)}$ satisfies a linear stochastic equation, thus it can be written as a sum
over space-time paths and can be interpreted as the canonical partition sum of a directed polymer (DP)
with endpoints $(0,0)$ and $(x,t)$ in a unit white noise random potential $-\eta$ [see Fig.~\ref{fig-polymers}(a)]
\be \label{zdef} 
Z(x,t|y,0) = \int_{x(0)=0}^{x(t)=x}  Dx e^{-  \int_0^t d\tau [ \frac{1}{4}  (\frac{d x}{d\tau})^2  - \sqrt{2} ~ \eta(x(\tau),\tau) ]}.
\ee
 The function $P_\Delta(\sigma_1,\sigma)$ maps to the
JPDF of the free energies of two DP starting both in $(0,0)$ but ending in $(0,t_1)$ and 
$(0,t_2=t_1(1+\Delta))$: in Fig.~\ref{fig-polymers}(a) is shown a typical configuration of these two paths, which
tend to visit the lower valleys of the potential (bluer regions), i.e. faster growth regions, compatible
with their boundary conditions and kinetic energies, which tend to minimize their length.
The JPDF \pref{JPDF0} is obtained \cite{SuppMatt}
from the joint integer moments 
$\langle { Z(0,t_1)^{n_1} Z(0,t_2)^{n_2} } \rangle$, averaged over realizations
of $\eta$. They are given by the quantum mechanical amplitude of the following process [see Fig.~\ref{fig-polymers}(b)]: $n_1 + n_2$ 
bosons with pair-wise attractive potential evolve from $x=0$ in imaginary time
up to time $t_1$. At $t= t_1$, $n_1$ of them are annihilated at $x=0$, 
while the other $n_2$ keep evolving up to $t= t_1(1 + \Delta)$, { at which} they are all finally destroyed at $x=0$ (Eq.(S3) in \cite{SuppMatt}). 
Their Hamiltonian is the Lieb-Liniger Hamiltonian 
with attractive interaction
\begin{equation}\label{LLH}
H^{(n)}_{LL}  = -\sum_{j=1}^n \partial^2_{x_j} -2 \sum_{i<j}^n \delta(x_i - x_j)  - \frac{n}{12}
\end{equation}
\begin{figure*}[t]
 \centering
 \includegraphics[width=\hsize]{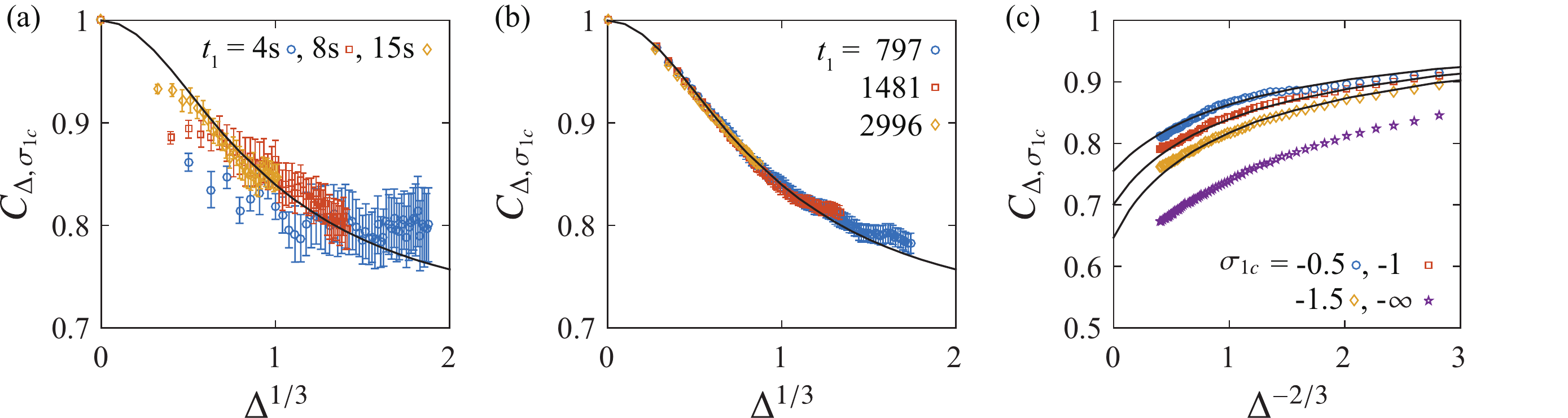}
  \caption{
The conditional covariance $C_{\Delta,\sigma_{1c}} = C(t_1,t_2)_{\tilde{h}_1> \sigma_{1c}} / C(t_1,t_1)_{\tilde{h}_1> \sigma_{1c}}$ (\eqref{CorrDef}). 
\textbf{a},\textbf{b}, Experimental (a) and numerical (b) results for $\sigma_{1c} = -1$ and with varying $t_1$ (symbols), compared with the theoretical prediction (black line). The error bars indicate the standard errors. To reduce the effect of finite-time corrections, here we used such realizations that satisfy $\tilde{h}_1 > \tilde{h}_{1c}$ with $\text{Prob}[\tilde{h}_1 \geq \tilde{h}_{1c}] = 1-F_2(\sigma_{1c})$.
\textbf{c}, Numerical data for $t_1 = 1008$ and for $\sigma_{1c} = -0.5, -1, -1.5$ and $-\infty$ (unconditioned). Error bars are omitted here for the sake of visibility. The black lines indicate the theoretical predictions for finite $\sigma_{1c}$. { At large $\Delta$ and for any $\sigma_{1c} $ they converge to their asymptotic values as $C_{{\Delta \to \infty,\sigma_{1c}}} + A_{\sigma_{1c}}\Delta^{-2/3} + B_{\sigma_{1c}}\Delta^{-1} + \ldots$}. For $\sigma_{1c} = -\infty$ (the unconditioned case), the theory suggests a strictly positive asymptotic value, specifically $C_{\infty, -\infty} \approx 0.6$, which is consistent with the trend of the unconditioned data set in the panel \textbf{c} (purple stars).
}
  \label{fig-corr}
\end{figure*}%
extensively studied recently in the context of integrable out-of-equilibrium dynamics \cite{cc-07,Panfil,Piroli}. 
Integrability of this dual quantum model
allows us to derive an analytical expression for $P_{\Delta}(\sigma_1, \sigma)$,
in the form $P_{\Delta}(\sigma_1, \sigma) = P^{(1)}_{\Delta}(\sigma_1, \sigma) ( 1 + \mathcal{O}(e^{- \frac{4}{3} \sigma_1^{3/2}}))$ with $P^{(1)}_{\Delta}(\sigma_1, \sigma)$ exactly determined in this work 
(see \cite{SuppMatt} for more details).
It is written as a trace of kernels acting on $\mathbb{R} \times \mathbb{R}$:
\begin{align}\label{pdfresult}
&P^{(1)}_{\Delta}(\sigma_1, \sigma) =  \left( \partial_{\sigma_1} \partial_{\sigma} -  {\Delta^{- 1/3}} \partial^{2}_{\sigma} \right) \times  \\ & \Big\{ F_2(\sigma)     \text{\text{Tr}} \Big[ \Delta^{1/3} \Pi_\sigma K^{\Delta}_{\sigma_1}   \Pi_\sigma   (I-\Pi_\sigma K_{\Ai}\Pi_\sigma )^{-1}   - \Pi_{\sigma_1} K_{\Ai} \Big] \Big\}. \nonumber
\end{align}
where $\Pi_\sigma$ projects on the interval $[\sigma,+\infty] \in \mathbb{R}$, $I$ is the identity operator
and $\partial_\sigma$ denotes partial derivatives. The expression involves 
the well-known Airy kernel  $K_{\Ai}(r,r')= \int_0^\infty dz \Ai(r + z) \Ai(r' + z)$  
from random matrix theory \cite{TW1994} 
and a novel kernel 
 \begin{align}\label{eq:Kfinal}
 K^{\Delta}_{\sigma_1}(r,r')  & = \int_0^\infty dz_1 dz_2 \Ai(-z_1 + r)  \Ai(-z_2 + r')    \nn \\& \times   K_{\Ai} (z_1 \Delta^{1/3} + \sigma_1, z_2 \Delta^{1/3} + \sigma_1).
 \end{align}
 The formula \pref{pdfresult} 
is simple enough 
for numerical evaluations  
for any value of $\sigma_1$ inside its 
expected validity range (specifically $\sigma_1 \gtrsim  -1.5$). 
This allows us to perform 
direct tests of the theoretical predictions, both experimentally and numerically, without any fitting parameters. 
Experimentally, we study growing interfaces of { electrically-driven} liquid-crystal turbulence, which were previously shown to be in the KPZ class \cite{TS2010,TSSS2011,TS2012} { (see Supplemental Material \cite{SuppMatt} and \cite{TS2012} for details).
We use 955 interfaces, generated from a turbulent nucleus (droplet initial condition) triggered by laser, 
for which the non-universal parameters $v_\infty$ and $\Gamma$
were determined with high precision \cite{TS2012}
 and used to obtain the rescaled variables.
Then we measure}
the conditional cumulants $\cum{\tilde{h}_{12}^n}{\tilde{h}_1> \sigma_{1c}}$
 { with different $t_1$  [Fig.~\ref{fig-condcum}(a)]}. 
Their asymptotic forms, { which are} indicated by overlapping of data sets, { are found to show} an excellent agreement with the theoretical predictions [Figs.~\ref{fig-condcum}(a) and S3(a) in \cite{SuppMatt}]. 
{ For a further test, we carry out} numerical simulations of the off-lattice Eden model \cite{T2012} { (5000 realizations; see \cite{SuppMatt} for details) and the same quality of agreement is obtained} [Figs.~\ref{fig-condcum}(b) and S3(b) in \cite{SuppMatt}]. 
We also measure the conditional covariance \pref{CorrDef} and find agreement both experimentally and numerically (Fig.~\ref{fig-corr}). This 
indicates that our predictions
describe universal time correlation of the droplet KPZ interfaces.

\begin{figure}[t]
 \centering
 \includegraphics[width=\hsize]{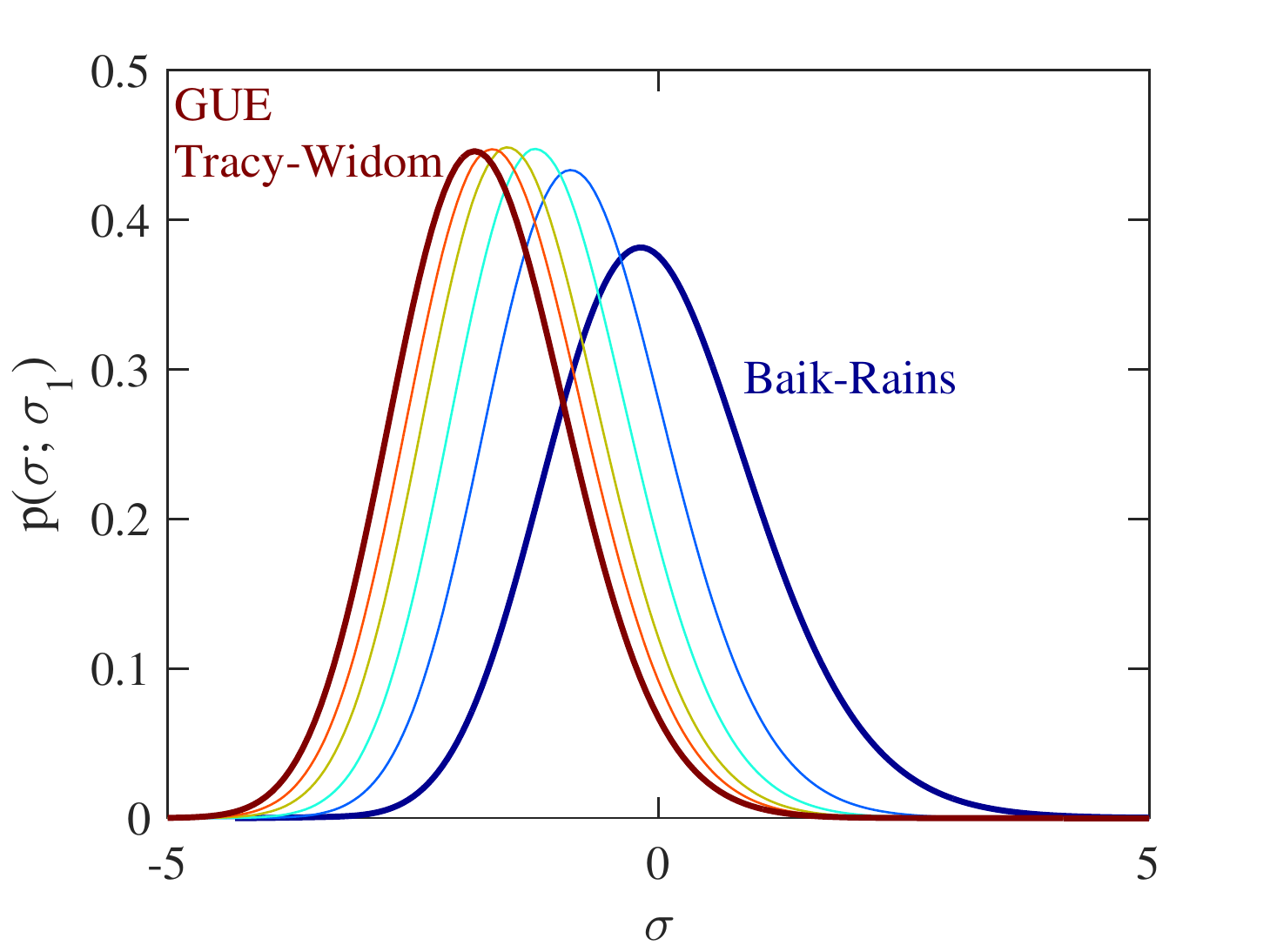}
  \caption{{\blue Predicted crossover from the Baik-Rains distribution ($\Delta \to 0$, i.e. $t_2/t_1 \to 1^+$) to the GUE Tracy-Widom distribution ($\Delta \to \infty$, i.e. $t_2/t_1 \to \infty$),
for the conditional probability of the rescaled height difference $\tilde h_{12}$. Plotted here is the normalized conditional probability $P^{(1)}_\Delta(\sigma_1=0, \sigma)/F_2'(\sigma_1=0)$ with $\Delta^{1/3} = 0, 0.7, 1.4, 2.8, 5.6, \infty$ from right to left.}
}
  \label{fig-crossover}
\end{figure}

Moreover,
our theory shows analytically for the first time the crossover between different probability distributions for 
$\tilde{h}_{12}$, as $\Delta$ varies (see Fig.~\ref{fig-crossover}). In the limit of close times, 
$t_2/t_1 \to 1^+$ 
the JPDF \pref{pdfresult} factorizes 
and $\tilde{h}_{1}, \tilde{h}_{12}$ become two independent random variables
following respectively the GUE Tracy-Widom and Baik-Rains distributions.
The emergence of the Baik-Rains distribution
is direct evidence of the approach to the KPZ stationary state when $t_2/t_1 \to 1^+$
{}\cite{FerrariSpohn2times}. 
As time separation increases,
a non-trivial aging form develops and 
for $\Delta \to \infty$ the joint statistics factorizes into the product of two GUE Tracy-Widom distributions. 
The next order correction, of order $\mathcal{O}(\Delta^{-1/3})$, gives access to the asymptotic value of the persistent correlation $C_{\Delta\to\infty, \sigma_{1c}}$: as $\sigma_{1c}$ decreases from $+\infty$ to $-\infty$,
i.e., the unconditioned case,
it is predicted to decrease from $1$ to a strictly positive value 
estimated to be $\approx 0.6$ {}\cite{SuppMatt}, which is consistent with our numerical data [Fig.~\ref{fig-corr}(c), purple stars].
The directed polymer/growth-history path representation enlightens 
this ergodicity breaking phenomenon. As shown in Fig.~\ref{fig-polymers}, the two polymers 
tend to visit the same minima of the random potential (corresponding to height fluctuations with large and positive rescaled amplitudes), thus sharing a finite fraction of their paths (growth histories) 
in the time interval $[0,t_1]$, see Fig. \ref{fig-polymers}.
This translates into a finite two-time correlation even in the limit 
$t_2 \gg t_1$, which our theory quantifies: as $\sigma_{1c}$ is increased, noise realizations with
large and positive fluctuations are selected in the interval $0 \leq t \leq t_1$, and therefore the shared path of the two polymers approaches unity and memory becomes perfect $C_{\Delta\to\infty, \sigma_{1c}} \to 1$.
This is consistent with the experimental and numerical observation (Fig.~\ref{fig-corr}). \\

\paragraph*{Conclusions.}
In summary, our results represent the first analytical theoretical predictions on the universal aging form 
for the two-time correlations of the KPZ equation \eqref{kpzeq}, 
{\blue that remarkably fit experimental and numerical data}. { \blue It
therefore gives a quantitative prediction for the crossover of the distribution of the fluctuations $\tilde{h}_{12}$ to the 
stationary state (i.e. the Baik-Rains distribution) as $t_2/t_1 \to 1^+$,
and confirms for the first time the breaking of ergodicity in the KPZ time evolution from the droplet initial condition. Both are proven to be universal properties 
shared by all growth processes in the KPZ class.  This universality in multi-time correlations, accompanied with ergodicity breaking could be explored in a broader class of growth problems
both within and beyond the KPZ class \cite{MBE1}. }
In expanding geometries, we expect similar persistence of memory 
when the spatial scale of dynamical correlations, $x \sim t_1^{\zeta}$ ($\zeta=2/3$ for KPZ) 
grows slower than the expanding substrate radius (here $\sim t_1$). 
It should also be relevant 
for other non-equilibrium systems, %
such as driven Bose-Einstein condensates \cite{AltmanBEC-KPZ} and
genetic segregation in expanding bacterial colonies \cite{nelson},
both shown to relate to KPZ. \\



We thank { P. Calabrese, I. Corwin,} and K. Johansson for discussions.
This work is supported in part by KAKENHI from JSPS, Nos. JP25103004, 16H04033, 16K13846 (K.A.T.), LabEX ENS-ICFP:ANR-10-LABX-0010/ANR-10-IDEX-0001-02 PSL* (J.D.N.), and the National Science Foundation under Grant No. NSF PHY11-25915.

{




\onecolumngrid

\begin{center}
{\large{\bf Supporting Information for\\ ``Memory and universality in interface growth''}}
\end{center}

\subsection{The KPZ equation and the two-time JPDF via the directed polymer}\label{meth:two-time}

To solve the KPZ equation, we use the Cole-Hopf mapping \cite{kardareplica}, which, given an initial condition 
$h(x,0)$, allows to write the height at time $t$ as  
$e^{h(x,t)} = Z(x,t) \equiv  \int dy Z(x,t|y,0) e^{h(y,0)}$. Here 
$Z (x,t|y,0)$ is the partition function of a continuum elastic polymer parametrized by the function $x(\tau)$ with $\tau \in [0,t]$ (its elastic energy given by $\frac{1}{4}  (\frac{d x(\tau)}{d\tau})^2$) in the random potential
$- \sqrt{2} ~ \eta(x,t)$ with fixed endpoints at $(x,t)$ and $(0,0)$:
\be \label{zdef} 
Z(x,t|y,0) = \int_{x(0)=0}^{x(t)=x}  Dx e^{-  \int_0^t d\tau [ \frac{1}{4}  (\frac{d x}{d\tau})^2  - \sqrt{2} ~ \eta(x(\tau),\tau) ]}
\ee
{ with $\overline{\eta(x,t) \eta(x',t')}=\delta(x-x') \delta(t-t')$.}
We focus on the droplet initial condition $h_{w_0}(x,0) =  - w_0  |x|$ in the limit $w_0 \to + \infty$.  Since $h(x,t)= \ln Z (x,t|0,0) $, the droplet height profile $h(x,t)$ corresponds to the (minus) free energy of a DP (at unit temperature) in a random potential,
going from the point $(0,0)$ to $(x,t)$.

To obtain the JPDF we introduce the generating function
\be \label{gener} 
g_{\Delta,  t_1 }(\sigma_1, \sigma)= \left\langle \exp \left( - e^{- t_1^{1/3} ( \tilde{h}_{t_1} - \sigma_1) - t_2^{1/3} ( \tilde{h}_{t_2} - ( \sigma_1 +  \Delta^{1/3} \sigma))  }\right) \right\rangle 
\ee
In the limit $t_1,t_2 \to +\infty$ it becomes equal to $g_{\Delta}(\sigma_1, \sigma)$ which yields the desired JPDF as
$P_{\Delta}(\sigma_1, \sigma) =( (\partial_{\sigma_1} \partial_\sigma - \Delta^{-1/3} \partial^2_{\sigma})  g_{\Delta}(\sigma_1, \sigma))$. On the other hand the first exponential in \eqref{gener} can be expanded 
as a series in positive integer powers of $e^{t_1^{1/3} \tilde{h}_{t_1}} = Z(0,t_1)$ 
and $e^{t_2^{1/3} \tilde{h}_{t_2}} = Z(0,t_2)$. The necessary ingredients are then the joint moments $\langle { Z(0,t_1)^{n_1} Z(0,t_2)^{n_2} } \rangle$. After integration over the Gaussian white noise $\eta$,
using replica techniques introduced by Kardar \cite{kardareplica}, they can be expressed in terms of quantum mechanical amplitudes
for a problem of attractive bosons in one dimension 
\bea \label{eq:partitionfunctionbosons}
&& \langle { Z(0,t_1)^{n_1} Z(0,t_2)^{n_2} } \rangle  = \prod_{j=1}^{n_2}\int_{-\infty}^{\infty} dy_j   \ { \langle \vec{0}_{n_2}
 }_{ }   | e^{-  (t_1 \Delta) H^{(n_2)}_{LL}} | \vec{y}_{n_2}  \rangle   \langle \vec{y}_{n_2}, \vec{0}_{n_1} | e^{- t_1 H^{(n_2+n_1)}_{LL} } | \vec{0}_{n_2}, \vec{0}_{n_1} \rangle 
\eea
where $| \vec{y}_{n_2}, \vec{x}_{n_1} \rangle$ denotes the quantum state where 
$n_2$ bosons are in positions $\vec{y} = \{ y_j\}_{j=1}^{n_2}$ and
$n_1$ bosons are in positions $\vec{x} = \{ x_j\}_{j=1}^{n_1}$ (see Fig.~2 in the main text). Here $H_{LL}^n$ is the Lieb-Liniger Hamiltonian for $n$ attractive bosons interacting with a point-wise potential 
\begin{equation}\label{LLH}
H^{(n)}_{LL}  = -\sum_{j=1}^n \partial^2_{x_j} -2 \sum_{i<j}^n \delta(x_i - x_j)  - \frac{n}{12}
\end{equation} \\
This Hamiltonian is integrable and all its eigenstates with $n$ particles, $ | \mu_n \rangle $, and eigenvalues, $E_{\mu_n}$, are explicitly known. We then expand \eqref{eq:partitionfunctionbosons} on the complete basis inserting $1 = \sum_{\mu_n} | \mu_n \rangle \langle \mu_n | $ in each factor. This leads to an expression
for each joint moment. The full summation over all eigenstates and $n_1,n_2$ is beyond reach
at present, but we are able to find an analytical expression for a partial summation 
when the second factor in \eqref{eq:partitionfunctionbosons} is restricted to the ground
state (for more details see section \ref{suppmatt:LL}). This already contains important information, since it
leads to $g^{(1)}_\Delta(\sigma_1, \sigma)$, equal to $g_\Delta(\sigma_1, \sigma)$ 
up to super-exponentially small corrections when $\sigma_1$ is positive. 
Finally we then obtain the JPDF
with the Kernel $ K^{\Delta}_{\sigma_1}$. 

\subsection{Summation over eigenstates of the equivalent bosonic problem and the two-time JPDF}\label{suppmatt:LL}

Here we give the explicit expression of the joint moments in terms of eigenstates of the Lieb-Liniger Hamiltonian
$H^{(n)}_{LL}$ \eqref{LLH}
 and we discuss the final steps leading to our result for the JPDF.
After inserting two complete sets of eigenstates with different particle numbers, $| \mu_{n_2} \rangle$ in the
first and $| \gamma_{n_1+n_2} \rangle$ in the second factor of equation \eqref{eq:partitionfunctionbosons}, we
obtain for the joint moments
 \begin{align}\label{moments2}
\langle { Z(0,t_1)^{n_1} Z(0,t_2)^{n_2} } \rangle & =  \sum_{\mu_{n_2}}  \sum_{\gamma_{n_1+n_2}}
 { \langle \mu_{n_2}  |  \vec{0}_{n_2} \rangle  \langle \vec{0}_{n_1 + n_2} | \gamma_{n_1 + n_2 } \rangle }{ } e^{- (t_1 \Delta) E_\mu - t_1 E_\gamma } F^{n_2;n_1+n_2}_{\mu ; \gamma} 
\end{align}
 where $\langle \mu_n | \vec{0}_{n} \rangle$ is the normalized overlap between the eigenstate $| \mu_n \rangle$ and the state $| \vec{0}_{n} \rangle$ where all the particles are in the same position $x=0$.  The quantities $F^{n_2;n_1+n_2}_{\mu ; \gamma}$ are the so-called form factors related to the matrix elements of powers of the bosonic annihilation operator between the two eigenstates. In the Lieb-Liniger model with $n$ particles,
the eigenstates are parametrized by a set of distinct (in general complex) quasi-momenta or rapidities
$\mu \equiv \{ \lambda_1,..\lambda_n\}$
and the eigenenergies are given by the sum 
$E_\mu=\sum_{\alpha=1}^n \lambda_\alpha^2 - \frac{n}{12}$. Focusing on the limit of
infinite system size $L$,
the general eigenstate is built by partitioning the $n$ particles into a set of $n_s \leq n$ bound states called { strings} \cite{cc-07}
formed by $m_j \geq 1$ particles with $n=\sum_{j=1}^{n_s} m_j$. 
Their associated rapidities are parameterized as 
$\l_{j, a_j}=k_j +\frac{i}2(m_j+1-2a_j)$,
where $m_j k_j \in \mathbb{R}$ is the total momentum of string $j$. 
Here, $a_j = 1,...,m_j$ labels the rapidities within the string $j=1,\dots n_s$. 
From now on a string state is denoted as $|\mu \rangle = |{\bf k}, {\bf m} \rangle$
and labeled by the set of $(k_j,m_j)_{j=1,..n_s}$. Its associated eigenenergy is thus
$E({\bf k},{\bf m}) \equiv  \sum_{j=1}^{n_s} m_j k_j^2-\frac{1}{12} m_j^3$,
which is the sum of a kinetic and binding energy. \\

In the infinite system size limit, the coupled sum over quasi-momenta$ \sum_{\{ k_j\}_{j=1}^{n_s}}$ can 
be replaced by a product of integrals, a property also used in \cite{we-narrow,dotsenko,KPZFlat,KPZStat}
\begin{equation}
\sum_{\mu_n} | \mu_n \rangle \langle \mu_n |  \xrightarrow[L \to \infty]{} \sum_{n_s=1}^n \frac{1}{n_s!} \left(\prod_{j=1}^{n_s} \sum_{ m_j =1}^{n} \int_{-\infty}^{+\infty} \frac{d k_j  m_j }{2 \pi} L \right) |{\bf k}, {\bf m} \rangle \langle {\bf k}, {\bf m}| \   \delta_{ \sum_{j=1}^{n_s} m_j , n   } 
\end{equation}
where the combinatorial factor $\frac{1}{n_s!}$ avoids double counting of states. 
Using that $\langle {\bf k}, {\bf m}|  \vec 0_{n} \rangle = L^{-n_s}  \Phi[\boldsymbol{k}, \boldsymbol{m}] \prod_{j=1}^{n_s} m_j^{-2} $  where $ \Phi[\boldsymbol{k}, \boldsymbol{m}] = \prod_{1\leq i<j\leq n_s} 
\frac{4(k_i-k_j)^2 +(m_i-m_j)^2}{4(k_i-k_j)^2 +(m_i+m_j)^2}$ \cite{we-narrow,dotsenko}
we can rewrite the sum in \eqref{moments2} as 
\bea 
  &&\langle { Z(0,t_1)^{n_1} Z(0,t_2)^{n_2} } \rangle   =  \sum_{n^{\mu}_s=1}^{n_2}  \sum_{n^{\gamma}_s=1}^{n_1+n_2}   
\frac{1}{n^{\mu}_s! n^{\gamma}_s!}  \prod_{j=1}^{n_s^\mu} \left(\sum_{m^\mu_j =1}^\infty \int \frac{dp_j}{2 \pi m_j^\mu }  \right)  \prod_{j=1}^{n_s^\gamma} \left(\sum_{m^\gamma_j =1}^\infty  \int \frac{dq_j}{2 \pi m_j^\gamma } 
 \right) F^{n_2;n_1+n_2}_{{\bf p} , {\bf m}^\mu;{\bf q} , {\bf m}^\gamma } 
 \nn  \\
  && 
  \times    \delta_{n_2, \sum_{j=1}^{n_s^\mu} m_j}
\delta_{n_1+n_2,\sum_{j=1}^{n_s^\gamma} m_j^\gamma}  \Phi[\boldsymbol{p}, \boldsymbol{m}^\mu]  \Phi[\boldsymbol{q}, \boldsymbol{m}^\gamma] 
e^{- (t_1\Delta )  E({\bf p}, {\bf m}^\mu) - t_1 E({\bf q}, {\bf m}^\gamma) } \nn \\
  && \label{momform1} 
\eea
where the superscripts $\mu$ and $\gamma$ indicate which set of string eigenstate is considered. 
Since the general case is out of reach at present, in this work we restrict to the case where the state $| \gamma_{n_1 + n_2} \rangle$ is constituted only by one single string of size $m = n_1 + n_2$ and momentum $q$ (the case $q=0$ corresponds to the ground state of the Hamiltonian \eqref{LLH}, where all the particles form a single bound state with zero momentum). The validity of this approximation
is discussed in Section \ref{suppmatt:tail}.
In this case $n_s^\gamma=1$ and the form factors 
can be expressed as
\begin{align} \label{ffformula} 
& F^{n_2;n_1+n_2}_{{\bf p} , {\bf m};q , n_1+n_2 } =  n_2! (n_1+n_2)! (  n_1)_{n_2}
\prod_{j=1}^{n_s} \frac{1}{(0^+ + i (q - p_j) + \frac{n_1+n_2-m_j}{2})_{m_j} 
(0^+ - i (q - p_j) + \frac{n_1+n_2-m_j}{2})_{m_j}}  
\end{align}
where $0^+$ is a regulator that is set to zero in the final stages of the calculation and $(n)_m$ are the Pochhammer symbols. It turns out that the sum in Eq. \eqref{momform1}, when 
restricted to the case $n_s^\gamma = 1 $, i.e. over the remaining string lengths $m_j^\mu$ and moments $p_j$,
can be explicitly performed for arbitrary $t_1$, using extensions of methods introduced 
in \cite{we-narrow,dotsenko,KPZFlat,KPZStat}. Taking the limit $t_1 \to +\infty$, we
obtain the generating function $g_\Delta^{(1)}(\sigma_1,\sigma)$ discussed in section \ref{meth:two-time}.
This leads to 
the final result for the two-time JPDF (Equation 4 in the text)in the limit $t_1 \to \infty$. 

Finally we would like to point out that in Ref. \cite{dotsenko2times1} the sum over states
has been implicitly restricted as $n_s^\gamma \geq n_s^\mu$. 
Hence important terms, not vanishing in the large time limit, have been neglected. 

\subsection{Tail approximation} \label{suppmatt:tail}
\begin{figure}
\centering
\includegraphics[scale=1.3]{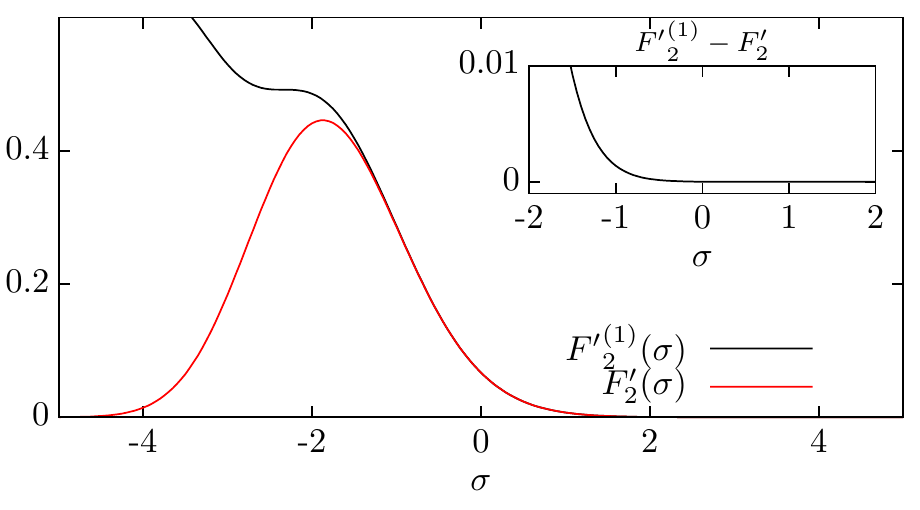}
\caption{Plot of the Tracy-Widom distribution $F'_2(\sigma)  \equiv \partial_\sigma F_2(\sigma)$ (red line) (with $F_2(\sigma) = \det [ 1- \Pi_\sigma K_{\Ai}]$) compared with its tail for positive $\sigma$ given by $F'{}_2^{(1)}(\sigma)$ (black line). Inset: difference between the Tracy-Widom distribution $F'_2(\sigma)$ and its tail $F'{}_2^{(1)}(\sigma)$, given in \eqref{sr2}.   }
\label{fig:gue_approx}
\end{figure}

In this section we discuss the validity of our approximation. The exact JPDF satisfies 
the sum rule
\bea
\int_{-\infty}^{+\infty} d\sigma P_\Delta(\sigma_1,\sigma) = F_2(\sigma_1) 
\eea
where $F_2(\sigma) = {\rm Det}[I - \Pi_\sigma K_{\Ai}]$ is the Tracy Widom 
cumulative distribution function associated to the GUE ensemble, which can be expressed
as a Fredholm determinant. On the other hand, our approximated result for the JPDF satisfies 
\bea \label{sr2} 
\int_{-\infty}^{+\infty} d\sigma P^{(1)}_\Delta(\sigma_1,\sigma) = F^{(1)}_2(\sigma_1) :=
1 - \text{\text{Tr}} [\Pi_\sigma K_{\Ai}] = 1 - \int_\sigma^{+\infty} dv K_{\rm Ai}(v,v) 
\eea
where we have defined
the "tail approximation" $F^{(1)}_2(\sigma)$ of the Tracy-Widom cumulative distribution.
It corresponds to keeping only the first term in the series expansion for a determinant of a matrix
near the identity ${\rm Det}[I +A] = 1 + {\rm Tr} A + \mathcal{O}(A^2)$. 
This function captures the leading (stretched) exponential behaviour for large and positive $\sigma$,  $F^{(1)}_2(\sigma) -1= \mathcal{O}(e^{- \frac{4}{3} \sigma^{3/2}})$ and 
the corrections are of higher (stretched) exponential order
$F_2(\sigma) = F^{(1)}_2(\sigma)  + \mathcal{O}(e^{- \frac{8}{3} \sigma^{3/2}})$. In fact,
one can see in Figure \ref{fig:gue_approx} that this approximation remains
excellent in a much broader range (with error less than $10^{-3}$ for any $\sigma > -1$). 
Hence, due to the sum rule (\ref{sr2}) we expect the same to be valid also for the accuracy of
our approximation $P^{(1)}_{\Delta}(\sigma_1, \sigma)$ of the JPDF as a function of $\sigma_1$.
Indeed, the experimental and numerical data for the conditional mean and variance are found to agree as well at $\sigma_{1c} = -1.5$ (Fig.~\ref{fig-condcum-suppl}).


\subsection{Two-time rescaled conditional covariance and its large $\Delta$ limit} \label{suppmat:corr}

The two-time rescaled conditional covariance, defined in the text in equation (3),
is calculated over the realizations
 such that $\tilde{ h}_{t_1} > \sigma_{1c}$. 
Using our formula for the JPDF and performing an expansion
at large $\Delta$ we obtain that the above covariance reaches a finite limit, given explicitly as
 \be \label{explicit}
 C_{\Delta\to\infty,\sigma_{1c}}:= \lim_{\Delta\to+\infty,\sigma_{1c}} C_{\Delta, \sigma_{1c}} =  1 + \frac{\int_{\sigma_{1c}}^{+\infty} d\sigma_1 \sigma_1 \tilde R_{1/3}(\sigma_1) 
- \mathcal{N}_{\sigma_{1c}}^{-1} \int_{\sigma_{1c}}^{+\infty} d\sigma_1 \sigma_1 F_2^{(1)}{}'(\sigma_1)  \int_{\sigma_{1c}}^{+\infty} d\sigma_1 \tilde R_{1/3}(\sigma_1)}{ 
\int_{\sigma_{1c}}^{+\infty} d\sigma_1 \sigma_1^2 F_2^{(1)}{}'(\sigma_1)  - \mathcal{N}_{\sigma_{1c}}^{-1} \left( \int_{\sigma_{1c}}^{+\infty} d\sigma_1 \sigma_1 F_2^{(1)}{}'(\sigma_1)   \right)^2}
 \ee
 where $\mathcal{N}_{\sigma_{1c}} = 1- F^{(1)}_2(\sigma_{1c})=\int_{\sigma_1}^\infty dy K_{\Ai}(y,y)$
 and where 
$\tilde R_{1/3}(\sigma_1)= \left[ \int_{\sigma_1}^\infty dy \Ai(y )\right]^2 - \int_{\sigma_1}^\infty dy K_{\Ai}(y,y)$.
The function $ C_{\Delta\to\infty,\sigma_{1c}}$ is plotted as a function of $\sigma_{1c}$ in 
Fig.~\ref{fig:coeff4}. It approaches $1$ with corrections of order $\sim \sigma_{1c}^{-1}$ for large and positive $\sigma_{1c}$. For $\sigma_{1c} < -1.5$ the tail approximation becomes unreliable (see Fig.~\ref{fig:gue_approx}), and the
minimum shown in the figure occurs at finite $\sigma_{1c}$, while the true minimum is expected
at $\sigma_{1c}= - \infty$. Nevertheless it suggests a rough estimate for the value of $C = \lim_{\sigma_{1c} \to - \infty} C_{\Delta\to\infty,\sigma_{1c}}$ as $ C_{\Delta\to\infty} \approx 0.6$.

\begin{figure}
\centering
\includegraphics[scale=1.2]{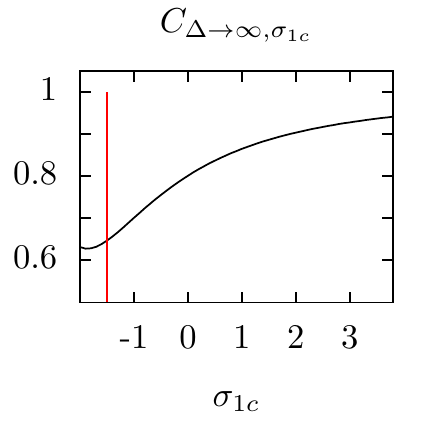}
\caption{
Two-time rescaled conditional covariance in the infinite $\Delta$ limit, defined in \eqref{explicit}, plotted as a function of $\sigma_{1c}$. The formula \eqref{explicit} is expected to give an excellent approximation for all $\sigma_{1c} \gtrsim -1.5$ (red vertical line), and deviate from the exact result
for smaller values. The curve approaches $1$ in the limit $\sigma_{1c} \to + \infty$. The minimum occurs at finite $\sigma_{1c} \approx -2$, while the true minimum (with value $C$) is expected at $\sigma_{1c}= - \infty$. The value at the minimum suggests that $C \approx 0.6$, consistent with the observations in the text (see Fig.~4 in the main text).}
\label{fig:coeff4}
\end{figure}

\subsection{Numerical evaluation { of analytical formulas}}\label{meth:numerical-ev}
For the numerical evaluation of traces and product of kernels and of the Tracy-Widom and Baik-Rains distributions ($F_2(\sigma)$ and $F_0(\sigma)$), we use a Gaussian quadrature discretization method to transform the continuous kernels into discrete square matrices. This amounts to approximate an integral with its discrete sum $\int_{s}^{\infty} dx f(x) \to \sum_{j=1}^n f(x_j) w_j$ where $\{ x_i\}_{i=1}^n $ are the $n$ quadrature points $x_j  \in [s, +\infty]$ and $\{ w_i\}_{i=1}^n$ the corresponding weights. Traces and Fredholm determinants become
standard traces and determinants of square matrices
\bea
 \text{Tr}[\Pi_s K] &\to& \text{Tr}_{i,j=1}^n[K(x_i,x_j) w_i] \\
 \det [ 1 - \Pi_s K] &\to& \det_{i,j=1}^n [\delta_{i,j} - w_j K(x_i,x_j) ]
\eea
and the inverse of the kernel $(I- \Pi_s K_{\Ai} \Pi_s)^{-1}$ is evaluated as the simple inverse of the 
corresponding discretized matrix.
The number $n$ 
depends on the particular quadrature rule 
and on the kernel but in general we find good convergence 
for $n \sim 50$ with a standard Gaussian quadrature rule (for the numerical evaluation of $F_2(\sigma) = \text{Det} [I  - \Pi_\sigma K_{\Ai}] $ exponentially fast convergence with $n$ was proved in \cite{Bornemann}). 

\subsection{Liquid crystal experiment}\label{meth:experiment}

For the experiment,
 we studied growing interfaces in electrically driven turbulence
 of nematic liquid crystal,
 which were previously shown to be in the KPZ class
 \cite{TS2010,TSSS2011,TS2012}.
The interfaces constitute boundaries between two distinct turbulent states,
 called the dynamic scattering modes (DSM) 1 and 2 in the literature,
 which are characterized by the absence and abundance, respectively,
 of sustained topological defects in the director field.
Under the experimental condition used here,
 DSM2 is stable while DSM1 is only metastable.
Therefore, by nucleating DSM2 locally amid DSM1,
 using ultraviolet laser pulses as a trigger,
 one can observe a growing cluster of DSM2, bordered
 by a fluctuating circular interface
 (as shown in Fig.1 of Letter).
We generated 955 such circular interfaces,
 for which the non-universal parameters $v_\infty$ and $\Gamma$ were
 previously determined with high precision \cite{TS2012}.
See Ref.~\cite{TS2012} for more detailed descriptions
 of the experimental system.

Note that, in such an isotropic system,
 the local radius of the circular cluster, measured in any angular direction,
 corresponds to the local height at the origin, $h(0,t)$,
 of the KPZ equation with the droplet initial condition.
This is because both measure
 the distance from the starting point of the growth
 along the mean growth direction.
Therefore, in the analysis we present in the Letter,
 we use data of the local radius at all angular positions
 to obtain statistically accurate results for $h(0,t)$.

\subsection{Eden numerics}\label{meth:eden}

The numerical test was carried out
 with an off-lattice version of the Eden model,
 which was introduced in Ref.~\cite{T2012}
 as a convenient model to study circular interfaces
 with isotropic growth speed.
The model consists of a cluster of round particles of unit diameter,
 added one by one,
 starting from a single particle placed in two-dimensional space.
At each time step, one attempts to add a new particle,
 next to a randomly chosen particle in a randomly chosen direction.
The attempt is adopted if it does not overlap with any existing particle,
 otherwise the attempt is simply withdrawn.
In either case, time is increased by $1/N$,
 where $N$ denotes the number of particles.
A circular interface is then obtained
 as the outmost perimeter of the cluster.
Here we generated 5000 independent realizations of such circular interfaces,
 using an efficient algorithm described in Ref.~\cite{T2012}.
The height function $h(x,t)$ is defined as in the experiment.
The values of $v_\infty$ and $\Gamma$ were taken
 from those from an extensive numerical study reported in Ref.~\cite{AOF2013}.
Since this system is also isotropic,
 similarly to the experiment,
 we use the local radius of the circular interfaces
 to compare with the theoretical predictions for $h(0,t)$.

\newpage
\section*{Supporting figures}

\begin{figure*}[h]
 \centering
 \includegraphics[width=0.5\hsize]{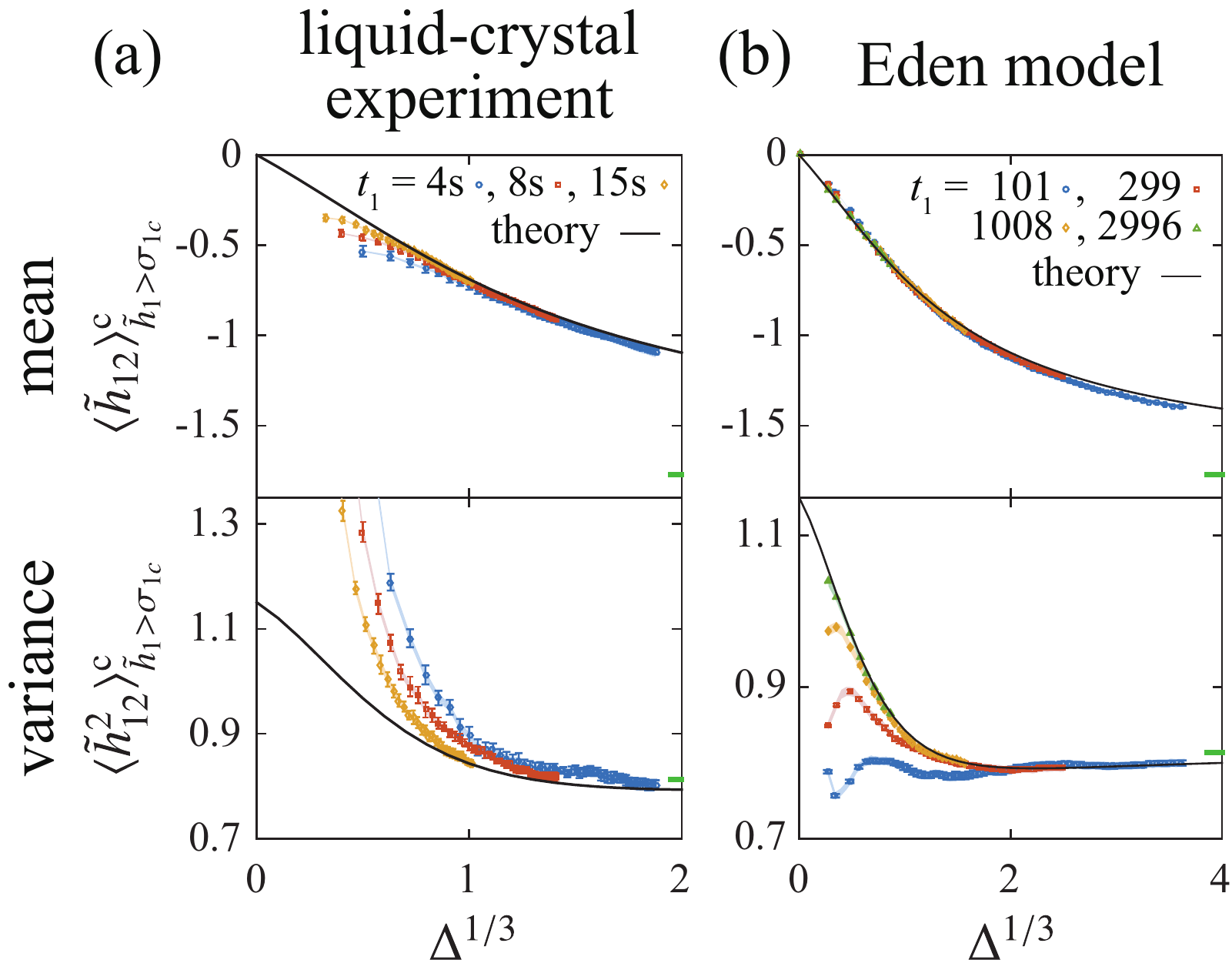}
  \caption{ 
Conditional mean $\cum{\tilde{h}_{12}}{\tilde{h}_1 > \sigma_{1c}}$ and variance $\cum{\tilde{h}_{12}^2}{\tilde{h}_1 > \sigma_{1c}}$ with $\sigma_{1c} = -1.5$. Experimental (a) and numerical (b) data with different $t_1$ are shown in different colors and symbols. The regions of overlapped data indicate the asymptotic $\Delta$-dependence, which is found to be in excellent agreement with the theoretical predictions (black lines), without any fitting parameter. The error bars indicate the standard errors, and the shaded areas show the uncertainty due to the estimation error in $v_\infty$ and $\Gamma$. To reduce the effect of finite-time corrections, here we used such realizations that satisfy $\tilde{h}_1 > \tilde{h}_{1c}$ with $\text{Prob}[\tilde{h}_1 \geq \tilde{h}_{1c}] = 1-F_2(\sigma_{1c})$. The deviation of the non-overlapped data is due to finite-time corrections, which decay as $t_1^{-1}$ (Fig.~\ref{fig-condcum-finitet}). Note that the asymptotic theoretical curves converge to the Baik-Rains values (mean 0, variance 1.1504) at $\Delta \to 0$ and to the GUE Tracy-Widom values (mean -1.7711, variance -0.8132, indicated in the figures by the green bars) at $\Delta\to\infty$.
}
  \label{fig-condcum-suppl}
\end{figure*}%

\begin{figure*}[h]
 \centering
 \includegraphics[width=.9\hsize]{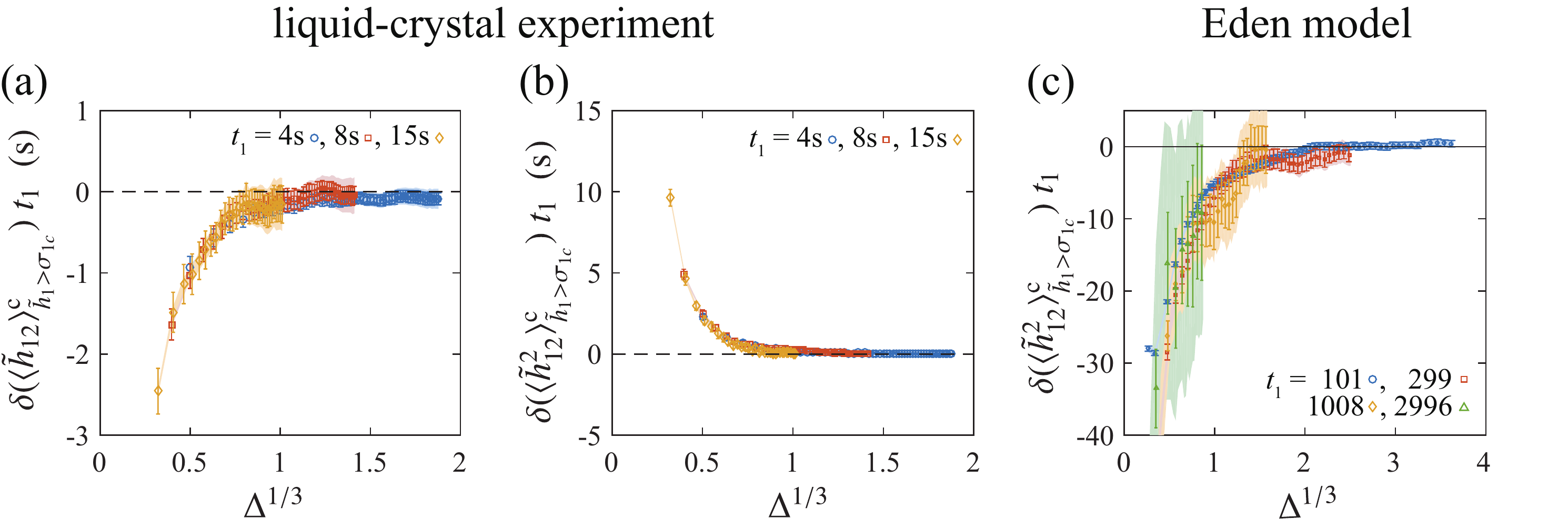}
  \caption{
Finite-time corrections in the conditional mean $\cum{\tilde{h}_{12}}{\tilde{h}_1 > \sigma_{1c}}$ and variance $\cum{\tilde{h}_{12}^2}{\tilde{h}_1 > \sigma_{1c}}$. Using the experimental (a,b) and numerical (c) data for $\sigma_{1c}=-1$ and different $t_1$, shown in Fig.~2 in the text, we plot the difference from the theoretically predicted values, $\delta(\cum{\tilde{h}_{12}^n}{\tilde{h}_1 > \sigma_{1c}})$, multiplied by $t_1$. Overlapping of the data indicates that the finite-time corrections decay as $\delta(\cum{\tilde{h}_{12}^n}{\tilde{h}_1 > \sigma_{1c}}) \sim t_1^{-1}$. The error bars indicate the standard errors, and the shaded areas show the uncertainty due to the estimation error in $v_\infty$ and $\Gamma$. We do not show the finite-time corrections in the conditional mean for the Eden-model simulation, which were too small to evaluate their $t_1$-dependence in a reliable manner.
}
  \label{fig-condcum-finitet}
\end{figure*}%

\newpage



\end{document}